Can Quantum Physics Help Fight Climate Change? A Thematic Analysis of Press Release Narratives

Maxwell Bernstein, Joanna Krajewski, and Rachel Fisher

School of Journalism and Mass Communication

The University of Iowa

**Author Note**

This research received no external funding.

The authors report no competing interests to declare.

The data supporting the findings of this study are publicly available through the NexisUni database.

ChatGPT was used to improve textual clarity for the reader, with rigorous editing from the authors. All research was done in accordance with the method listed in this paper.

The authors would like to thank Brian Ekdale for sharing a MAXQDA license.

Can Quantum Physics Help Fight Climate Change? A Thematic Analysis of Press Release Narratives

**Abstract**

The climate crisis demands innovative solutions and quantum technologies are considered by some a potential solution (Berger et al., 2021; Calvin et al., 2023). The United Nations declared 2025 the International Year of Quantum Science and Technology (United Nations, 2024), focusing on applications for quantum technologies. Quantum technologies, leveraging quantum mechanics, offer unprecedented computational power (Ajagekar & You, 2022) and sensitivity (Schleich et al., 2016) that could revolutionize various sectors, including climate action. However, an understanding of how discourse from academia, government, and industry frames quantum technologies as a climate solution is lacking (Suter, Ma, & Pöhlmann, 2024). This research aims to fill this gap by analyzing the narratives in press releases and helping inform a successful democratization of quantum technologies.

Can Quantum Physics Help Fight Climate Change? A Thematic Analysis of Press Release Narratives

**Introduction**

Human-caused climate change has already led to widespread and severe impacts on ecosystems, human health, and livelihoods. While adaptation and mitigation efforts are growing, the rapid development and deployment of low- and zero-emission technologies is critical to limit future risks and ensure sustainable development (Calvin et al., 2023). Quantum computing and artificial intelligence offer emerging strategies to accelerate the design of sustainable energy materials, optimize energy systems, and enable large-scale renewable energy integration, contributing to the global effort toward carbon neutrality. Specific applications include using quantum algorithms to simulate molecular systems for next-generation battery materials, deploying quantum machine learning to improve the forecasting and control of renewable energy sources, and applying quantum optimization to enhance the efficiency of energy supply chains and smart grids (Ajagekar & You, 2022). Complementing these advances, quantum sensors enable highly sensitive, interference-free measurements of environmental variables such as gravity, magnetic fields, and atmospheric changes, which could improve the detection of raw materials, aid in earthquake prediction, and allow for more accurate monitoring of sea-level rise due to climate change (Schleich et al., 2016).

**Literature Review**

These developments are part of what is often called the "Second Quantum Revolution," (Deutsch, 2020), which marks a shift from merely observing quantum phenomena to actively controlling, engineering, and applying them at the level of individual particles. Unlike the first quantum revolution, which brought technologies such as lasers, transistors, and MRI machines through an understanding of quantum mechanics at a collective level, the second revolution is

Can Quantum Physics Help Fight Climate Change? A Thematic Analysis of Press Release Narratives

centered on manipulating single quantum systems—ushering in new capabilities in quantum computing, communication, and sensing with profound technological and societal implications (Jaeger, 2018).

With the UN declaring 2025 as the International Year of Quantum Science and Technology (United Nations, 2024), attention is on quantum technologies and their applications. Capturing and sustaining this attention requires a nuanced understanding of how quantum technologies are framed across sectors. Research by Suter et al. (2023) shows that public narratives about quantum technology are shaped by five overarching themes: (1) technical aspects and applications, (2) politics and global conflicts, (3) people and society, (4) national technology strategies, and (5) business and market development. These themes reflect how quantum technologies are positioned—not only as breakthroughs in computation and sensing but also as instruments of geopolitical power, tools for societal progress, and opportunities for economic growth and commercialization (Suter et al., 2023).

Scant research currently exists on communication of quantum technologies and is especially lacking regarding how quantum technologies are positioned as a solution to climate change (Meinsma et al., 2023, 2025; Roberson et al., 2021, 2021, 2023; Suter et al., n.d.; Thomas et al., 2024; Vermaas, 2017; Wolbring, 2022).

Analyzing the communication surrounding quantum technologies reveals the importance of narrative and methodological approaches to deploying this technology in society. Storytelling plays a vital role in shaping public attitudes toward climate action and policy support. How storytelling surrounds the deployment of quantum technology, especially with climate action, will determine its impact on mitigating adverse impacts from climate change.

Can Quantum Physics Help Fight Climate Change? A Thematic Analysis of Press Release Narratives

## Methodology

In order to initiate scholarship on quantum technology communication and its application within climate change mitigation communication, this research employed the Constant Comparative Method of Qualitative Analysis (Glaser, 1965) to systematically examine emerging patterns and narratives within a sample of 100 press releases, selected from 305 documents collected via the NexisUni database.

### Sample

To obtain the press releases in our sample, press release documents in Nexis Uni were filtered using the search terms "quantum computer OR quantum technology OR quantum computing" AND "climate change OR global warming" and covered the period from January 1, 2024, to June 1, 2024.

### Analysis

The sample was coded and analyzed using MAXQDA software. Initially, codes were assigned inductively, meaning they were developed directly from the data without preconceived categories. As analysis progressed, coding became increasingly deductive, with new data compared against existing codes and themes. Following the core principle of the Constant Comparative Method (Glaser, 1965), each incident (or relevant article) was constantly compared with previous incidents coded under the same category. This iterative comparison allowed for the refinement of categories, identification of dimensions and properties of emerging themes, and development of an integrated, data-grounded understanding.

Throughout coding, thematic analysis was employed to extract key narratives, which were then categorized into ten overarching themes. In line with Glaser's (1965) methodological stages, this involved (1) comparing incidents applicable to each category, (2) integrating categories and their properties, (3) delimiting the theory by refining and reducing overlapping themes, and (4) articulating the final thematic structure in a coherent narrative form. Theoretical saturation was monitored, with coding continuing until additional data no longer provided new insights into the established categories. Sectoral differences were also examined by coding for the subject of each press release—academia, government, or industry—and analyzing how portrayals of quantum technologies' relevance to climate change varied across these domains. Findings were generated to remain closely tied to the press release content, ensuring that resulting narratives were plausible, integrated, and reflective of communication patterns observed in the data. In this analysis, over 8500 codes were applied to 100 articles, see Figure 1.

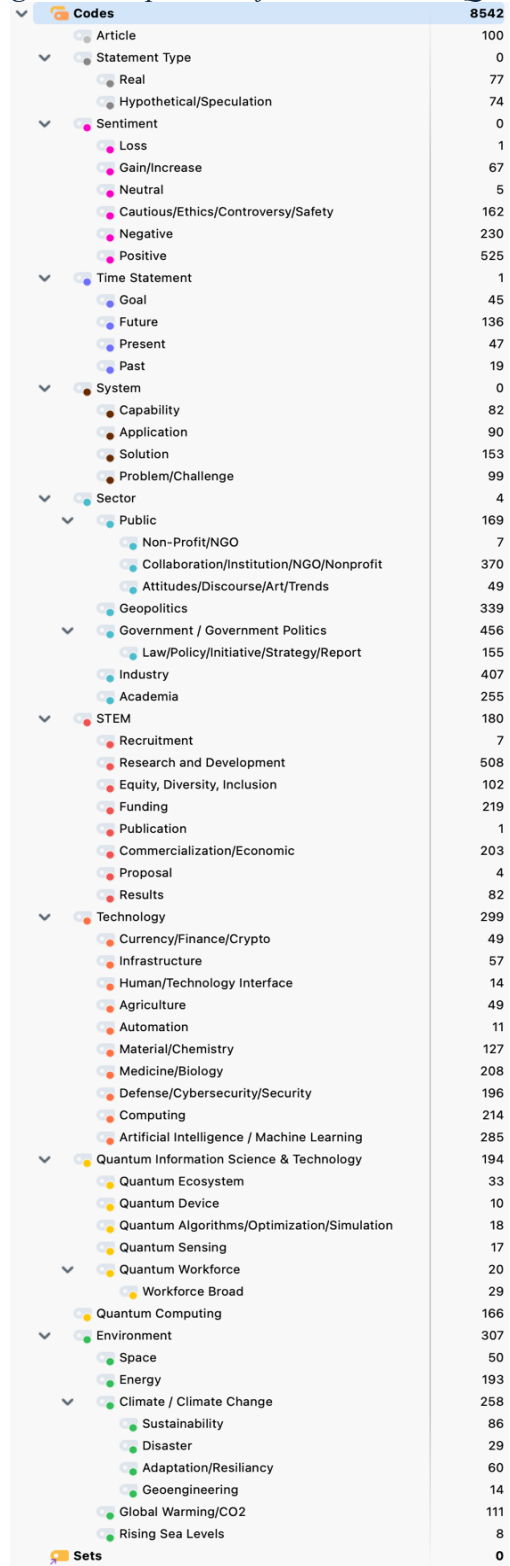

*Figure 1. Depiction of Codes in MAXQDA*

Can Quantum Physics Help Fight Climate Change? A Thematic Analysis of Press Release Narratives

## Results

This section presents the findings of a qualitative thematic analysis of press releases from academic, governmental, and industrial sources that position quantum technologies in relation to climate change. Quantum technologies are increasingly framed as vital tools for addressing climate change, from accelerating climate modeling and energy breakthroughs to enabling discovery of new materials. Often discussed alongside Artificial Intelligence (AI), quantum is part of a broader ecosystem of emerging tech aimed at environmental solutions. At the same time, press releases highlight cybersecurity and ethical risks, prompting calls for responsible development. Governments and industries are investing heavily in commercialization and national strategies, while universities drive research and talent pipelines. Equity, access, and global collaboration remain central to ensuring these innovations benefit all. See Table 1 below, which details the themes identified through this study.

*Table 1. Summary of Themes*

| Theme | Summary |
|---|---|
| **Quantum as a Climate Solution** | Quantum computing is framed as a transformative tool for improving climate modeling, discovering sustainable materials, and optimizing energy systems. This theme also includes the recognition of climate change as a threat. |
| **Synergy Between Quantum and Emerging Technologies** | Quantum technologies are frequently positioned alongside AI, biotechnology, digital twins, and other emerging technologies, suggesting integration will be key to achieving societal goals. |
| **Hope vs. Caution (Risks and Ethics)** | Narratives balance optimism with warnings about cybersecurity threats, ethical uncertainties, and societal impacts. |
| **Government Support and National Strategy** | Press releases highlight substantial government investment in quantum R&D for |
|  | economic competitiveness, climate resilience, and sovereignty. |
| **Industry Investment and Commercialization** | Firms emphasize real-world applications, from drug discovery to innovative energy solutions, showcasing pathways toward sustainability. |
| **Academic Leadership and Innovation Pipelines** | Universities are central to developing scalable architectures for research infrastructure and quantum technologies, workforce training, and foundational research. |
| **Equity, Access, Societal Inclusion, and Global Justice** | Efforts are being made to ensure broad, inclusive access to quantum technologies and to bridge potential future divides. |
| **Strategic Geopolitics and Sovereignty** | Quantum advancements are framed within geopolitical competition, emphasizing national security and influence. |
| **AI Governance and Ethical Technology** | Broader concerns about the responsible governance of emerging technologies are a recurring theme. |
| **Ecosystem Building** | Press releases stress the importance of collaborative ecosystems linking academia, government, and industry to scale quantum innovation. |

This analysis aimed to explore the way quantum technologies are framed—as innovations, as risks, and their role in broader socio-technological shifts, each reflecting a dominant narrative pattern. These themes collectively offer insight into how institutions seek to shape public understanding of quantum technologies' role in environmental futures. Below, each theme is discussed, with direct quotations from the press releases examined included, offering textual evidence for their narrative construction.

**Quantum as a Climate Solution**

A dominant theme across all sectors was the presentation of quantum computing as a transformative climate solution. In this framing, quantum systems are described as capable of Can Quantum Physics Help Fight Climate Change? A Thematic Analysis of Press Release Narrativessolving complex, computationally intractable problems, enabling new frontiers in sustainability research. One press release notes that quantum simulations might "determine how to reduce carbon dioxide in the atmosphere to help rein in climate change" (Agence France Presse – English, Scientists launch hub to channel quantum power for good, 05 March 2023). Another echoed this optimism, stating that the impacts of quantum computing will be far-reaching "with potential opportunities across multiple industries and advances in sectors such as healthcare, climate change, energy, communications, and finance" (Impact News Service, Can we build a safe and inclusive 'quantum economy'?, 06 February 2024). One of the boilerplates emphasized climate change as an application for quantum technologies, stating:

> Quantum computing has the potential to solve equations in mere minutes that would traditionally take existing computers tens of thousands of years to complete -- enabling immense potential to enhance medication discovery, improve artificial intelligence tools, combat climate change, and model complex scenarios (Targeted News Service, Calif. Gov. Newsom Announces World-Leading Science & Technology Research Center in Los Angeles, 4 January 2024).

These narratives underscore a belief in quantum's capacity to reimagine climate solutions.

**Synergy Between Quantum and Emerging Technologies**

Quantum technologies were frequently discussed with other emerging technologies, particularly AI, digital twins, and biotech. This narrative suggests that quantum will not act alone but will serve as a key enabler within hybrid technological ecosystems. For instance, one press release explained that "fostering synergy between AI and burgeoning quantum technologies is a vital strategy for steering AI towards sustainable development," citing quantum's ability to make

Can Quantum Physics Help Fight Climate Change? A Thematic Analysis of Press Release Narratives

AI models more compact with a reduced energy footprint (Africa Newswire, How to manage AI's energy demand - today, tomorrow and in the future, 25 April 2024). Another highlighted a partnership between Bosch and Multiverse Computing to deploy quantum-powered digital twins of factories for energy and waste optimization and quality control (MarketLine NewsWire (Formerly Datamonitor), Digital twin market to surpass $150bn by 2030, impacting many industries, 16 April 2024). The interdependency between quantum technologies and AI for ecological solutions can be exemplified by the following example:

> In the Global Big Ideas Competition that Science-i recently hosted, researchers worldwide submitted proposals for projects that would impact both their local forests and those on the other side of the globe. Incorporating everything from new artificial intelligence technologies to quantum computing to underground fungal networks that communicate with trees, these projects promise to add to humanity's shared knowledge base and help inform forest management decisions. (States News Service, Scientists from dozens of countries coming to Purdue for forestry collaboration in science-I bridging worlds workshop, 9 April 2024)

These examples point to an evolving discourse where quantum is not isolated but deeply interwoven with broader innovations.

**Hope vs. Caution: Risks and Ethics**

While many press releases emphasized quantum promises, several highlighted ethical and security concerns, painting a more nuanced picture. A common fear is quantum's potential to break modern encryption, enabling massive cybersecurity vulnerabilities. One article warned that quantum computing could be "a computer able to guess all possible password combinations at once" (States News Service, Entering the quantum era, 5 March 2024). These concerns were often placed in broader ethical discussions about the existential risks of emerging technologies. For example, one release noted that "quantum computing, AI, and geoengineering could pose existential threats to humanity if misused," raising the stakes for governance and precaution (ClimateWire, Her job: Ensuring AI and radical climate fixes do not backfire, 9 May 2024). The basis of this theme is a recognition of the powers within the technologies themselves.

**Government Support and National Strategy**

National investment and strategy were frequently highlighted as critical drivers for quantum innovation, often framed within broader economic and geopolitical ambitions. Press releases from Canada and the U.S. cited major funding programs supporting quantum R&D, with climate resilience positioned as a motivating factor. In British Columbia, for instance, over $11 million in funding was allocated to organizations "leading innovation in quantum computing" (States News Service, Government of Canada supports growth and innovation in British Columbia 's quantum technology industry, 30 May 2024), while Colorado introduced a Quantum Tax Credit bill projected to generate 10,000 jobs and fund university-industry collaboration (States News Service, State leaders announce details of bipartisan legislation to accelerate Colorado 's thriving quantum ecosystem, build on CU Boulder 's quantum legacy, 15 February 2024). Canada's national quantum strategy integrated climate-related sensing applications, particularly in the Arctic, as exemplified by the following:

> Over the next seven years, Canada will mobilize research teams with its allies to fine-tune the technology. It is currently in the first year of the strategy that calls for continuing research into quantum technology. The ensuing years will focus on the goals of Our

> North, Strong and Free. Quantum sensing has been specialized and applied to radar technologies that will be used to chart and monitor Arctic territory. Furthermore, with support from its Five Eyes allies, Canadian information is being made quantum-safe through extensive investment into cryptography (Canadian Press, Defence policy update focuses on quantum technology's role in making Canada safe, 1 May 2024).

These initiatives show how national strategies position quantum as a pillar of economic and military leadership.

**Industry Investment and Commercialization**

In industry press releases, the focus shifted to commercialization and real-world applications, especially sustainability-related ones. Companies emphasized modeling energy, healthcare, and environmental modeling breakthroughs, often through cross-sector collaborations. For example, startup Xatoms used quantum simulations to identify new molecules capable of purifying polluted water (ENP Newswire, Compute for Climate Fellowship Announces Inaugural Winners and Opens Applications for 2024 The program funds proof of concepts for new ideas leveraging advance...., 22 March 2024), while PsiQuantum and Mitsubishi worked on photoswitching molecules for applications such as solar energy storage and bright windows (ENP Newswire, -PsiQuantum, Mitsubishi UFJ Financial Group and Mitsubishi Chemical Announce Partnership to Design Energy-Efficient Materials on PsiQuantum's Fault-Tolerant Quantum Computer, 25 January 2024). The discourse here centered on quantum as a commercial driver for climate solutions, translating theoretical potential into applied innovation.

**Academic Leadership and Innovation Pipelines**

Can Quantum Physics Help Fight Climate Change? A Thematic Analysis of Press Release Narratives

Universities were consistently portrayed as the backbone of quantum innovation—responsible for foundational research and talent development. Oxford University researchers described the power of quantum computing through the analogy of the Bodleian library, "a classical computer would look through each book in turn to find the hidden golden ticket, potentially taking thousands of years; an advanced quantum computer could simply open every book at once." (States News Service, Entering the quantum era, 5 March 2024). Press releases emphasized universities' role in workforce development: "attract, develop, and retain talent to build a quantum-ready workforce" (States News Service, DPM Heng Swee Keat at the Asia tech x Singapore 2024 opening ceremony, 30 May 2024) and launching scholarship schemes to build a "quantum-ready workforce" (Impact News Service, Building an emerging technology pipeline in regional Australia, 25 January 2024). Press releases also mentioned establishing technology transfer programs in academia (Targeted News Service, MIT Tops Among Single-Campus Universities in U.S. Patents Granted, 11 April 2024). Academia was positioned not only as a research engine, but a social and economic pipeline for quantum readiness.

**Equity, Access, Societal Inclusion, and Global Justice**

A recurring concern in the dataset was ensuring equitable access to quantum technologies. Many press releases warned against repeating the "digital divide" and emphasized inclusive global development. As Jack Hidary of SandboxAQ stated, "We cannot accept that there will be a quantum divide" (Impact News Service, AI, and emerging technology at Davos 2024: 5 surprising things to know, 27 January 2024). Documents called for strategies that aim to broaden access to quantum technologies, such as:

Can Quantum Physics Help Fight Climate Change? A Thematic Analysis of Press Release Narratives

> The World Economic Forum's Quantum Economy Blueprint is a new guide for building a 'quantum economy' accessible to all nations and societies. This includes creating a national quantum strategy, broadening access to quantum technologies, and raising awareness about them. (Impact News Service, Can we build a safe and inclusive 'quantum economy'?, 06 February 2024)

Some also highlighted divides involving cybersecurity and access to technology, as the below demonstrates:

> As organizations race to adopt new technologies, such as generative AI, they should not lose sight of the risks created by well-flagged near-future applications of other technologies, such as quantum computing. Technological development is making the cyber equity gap stark within and between countries. This makes everyone more vulnerable, even the best-protected organizations and collaborative solutions that support those least able to secure themselves will benefit all. (Impact News Service, Global Risk Report 2024: The risks are growing, but so is our capacity to respond, 10 January 2024)

This theme underscores concerns around fairness, justice, harm reduction, and inclusivity in shaping the future quantum landscape.

**Strategic Geopolitics and Sovereignty**

Quantum was also framed within a geopolitical context, often as a race for technological supremacy. Press releases from China, Canada, and the U.S. emphasized the role of quantum in national sovereignty and strategic defense. For instance, Canada's use of quantum sensing to monitor Arctic territory, listed before, was presented as a national security imperative in the face of climate-driven polar access (Canadian Press, Defence policy update focuses on quantum

Can Quantum Physics Help Fight Climate Change? A Thematic Analysis of Press Release Narratives

technology's role in making Canada safe, 1 May 2024). Similarly, an academician of the Chinese Academy of Sciences said that China is the only country that has achieved the superiority of quantum computing. (Impact News Service, From space to the deep sea, China is at the forefront of exploring new fields - understanding the new advantages of China's economy, 8 April 2024). In these narratives, applications of quantum technologies were embedded within broader discussions of power and influence on the world stage.

**AI Governance and Ethical Technology**

The governance of AI and other emerging technologies appeared as a complementary theme, often intertwined with quantum. Policymakers and executives stressed the need for "agile governance" frameworks that adapt alongside fast-moving innovation (Impact News Service, AI, and emerging technology at Davos 2024: 5 surprising things to know, 27 January 2024). DARPA Director Prabhakar said that technology is "value-neutral"—ethical implications depend on how societies choose to wield it (ClimateWire, Her job: Ensuring AI and radical climate fixes do not backfire, 9 May 2024). Several releases emphasized the importance of integrating privacy, data governance, and equitable standards specifically into the deployment of AI. See an example of this, below:

> We have also strengthened the Emirate's capacity to use data by developing a data management and governance strategy to ensure its privacy and optimal use of advanced technologies such as AI and big data analytics (Emirates News Agency (WAM), Digital Readiness Retreat explores developing digital business models for government; launches State of Digital Transformation Report, 15 May 2024).

Another release noted Big Tech immense societal power and concerns around this regarding deploying emerging technology, stating:

> What we do know is that AI and other rapidly advancing technologies, such as quantum computing, biotechnology, neurotechnology, and climate-intervention tech, are becoming increasingly powerful and influential by the day. Despite the scandals and political and regulatory backlash of the past few years, Big Tech firms are still among the world's largest companies and continue to shape our lives in myriad ways, for better or worse (Impact News Service, Making emerging technologies safe for democracy, 28 March 2024).

These concerns reflect a broader effort to ensure technological advancement does not outpace ethical responsibility. The difference between this theme and *Hope vs. Caution: Risks and Ethics* involves managing, developing, and using technology to align with societal goals.

**Ecosystem Building**

Finally, ecosystem-building emerged as a key strategy to accelerate quantum innovation. This involved fostering collaboration between academia, government, and industry. One example was the NSF-funded Quantum Crossroads initiative led by the Chicago Quantum Exchange, which sought to connect top institutions and create scalable research models, as stated in the excerpt below:

> Beyond the 10 NSF Engines awards, a subset of the semifinalists and finalists were invited to pursue future NSF Engines development awards. A proposal led by the Chicago Quantum Exchange, an intellectual hub that connects top universities, national labs, and industry partners to advance the science and engineering of quantum information, train the future quantum workforce, and drive the quantum economy, could receive up to $1 million to further develop their partnerships and model for future NSF Engines projects. That proposal, Quantum Crossroads, was among the 16 finalists and the only one focused on quantum technologies (Business Wire, Innovate Illinois Celebrates NSF's Award to Propel Great Lakes ReNew and Advance Water Innovation, 2 February 2024)

Another example from Singapore described how coordinated partnerships between "data centre operators, enterprise users, and equipment suppliers" (States News Service, DPM Heng Swee Keat at the Asia tech x Singapore 2024 opening ceremony, 30 May 2024) to launch a new Tropical Data Centre standard to reduce emissions and improve energy efficiency for AI and data storage. These narratives emphasized that no single actor could deliver transformation alone; instead, the releases emphasize that systemic collaboration is vital for climate impact and innovation success.

## Discussion

There are several meaningful overlaps between the themes identified in Suter et al.'s (date) analysis and those found in this analysis of press releases on quantum technologies and climate change. Both studies highlight how technical innovation lies at the heart of the quantum discourse—Suter's (date) "Technical Aspects & Applications" aligns with the theme of "Quantum as a Climate Solution," which frames quantum technologies as essential tools for advancing climate modeling, improving materials for clean energy, and optimizing energy systems. Additionally, the frequent discussion of quantum technologies in conjunction with AI and other emerging technologies in this analysis corresponds with Suter et al.'s (date) inclusion of AI within broader technical applications, underscoring a shared narrative that hybrid technological ecosystems are key to addressing global challenges.

Beyond technical capabilities, both analyses emphasize the importance of broader societal and strategic contexts. Suter's (date) theme of "National Technology Strategies" corresponds with this studies findings under "Government Support and National Strategy," as both point to the strategic framing of quantum investments by governments in terms of economic growth, global leadership, and infrastructure development. Similarly, Suter's "People & Society" theme overlaps with this study's identified theme of "Equity, Access, Societal Inclusion & Global Justice" and "Academic Leadership and Innovation Pipelines." These shared themes reflect a growing recognition of the role of education, ethics, and inclusivity in of advancing quantum technologies. Together, the overlaps demonstrate how narratives also increasingly portray quantum technologies as technical innovations and as deeply embedded within societal, political, and ethical frameworks.

**Strategic Narratives and Sectoral Framing**

The findings of this study demonstrate that quantum technologies are increasingly framed as pivotal solutions to climate challenges, particularly in applications such as climate modeling, energy efficiency, and material discovery. Across academic, government, and industrial sectors, substantial attention is being devoted to investments in commercialization, infrastructure development, cross-sector collaborations, and workforce training—all aimed at accelerating the deployment of quantum technologies. Notably, quantum is often positioned alongside AI and biotechnology, suggesting that its potential may emerge not from isolated development but from integration with these complementary fields.

Nevertheless, the narrative surrounding quantum technologies extends beyond technical capabilities. Quantum is increasingly embedded within broader strategic frames tied to climate innovation, national competitiveness, and global leadership. This positioning reflects a growing perception of quantum not just as a scientific breakthrough but as a vital instrument for achieving societal, economic, and geopolitical goals. The prominence of cross-sector alliances for ecosystem building further illustrates this trend: partnerships such as those between Bosch and Multiverse and between PsiQuantum and Mitsubishi offer tangible examples of how quantum technologies are beginning to transition from theoretical promise to practical applications aimed at advancing sustainability. Similarly, academic institutions are portrayed as playing a crucial role in fostering commercial and workforce development.

**Equity, Ethics, and Inclusion in Quantum Narratives**

However, despite occasional acknowledgments of ethical, security, and access-related risks as identified in the results, these concerns often remain secondary to dominant narratives focused on innovation and investment. This imbalance may have profound implications, influencing public expectations, policy decisions, funding priorities, and regulatory frameworks. Left unchecked, it risks steering the development of quantum technologies toward a path that overlooks critical questions of equitable access, societal benefit, and unintended consequences. The strategic pairing of quantum with AI and other emerging technologies amplifies these dynamics, underscoring the urgency of ensuring that emerging technologies are developed inclusively and responsibly.

Can Quantum Physics Help Fight Climate Change? A Thematic Analysis of Press Release Narratives

**Managing Expectations and the Hype Helix**

Optimism about quantum technologies' potential contributions is widespread, as demonstrated by a sentiment analysis by Thomas et al. (2024). This is echoed in the results section where institutions describe quantum's ability to revolutionize sectors like climate change, yet simultaneously acknowledge potential ethical pitfalls, such as cybersecurity risks and misuse of power.

Yet, it must be contextualized within the broader dynamics of technological hype. Roberson et al. (2023) introduce the concept of the "hype helix," illustrating how national strategies amplified early expectations of quantum transformative capabilities. Over time, as quantum technologies progressed following the second quantum revolution in the early 2000s, these heightened expectations began cycling back to the research community through increased scrutiny, calls for ethical reflection, and demands for demonstrable societal benefits.

Additionally, Seskir et al. (2023) emphasize the importance of democratizing access to quantum technologies to ensure their benefits are broadly shared. While companies have expanded cloud-based access to quantum computers and educational resources—efforts often framed as democratization—these initiatives essentially serve instrumental goals aligned with market expansion and workforce development. True democratization, Seskir (2023) and his colleagues argue, requires deeper engagement with a broader range of stakeholders, including those indirectly affected by quantum technologies. It also demands a commitment to embedding these innovations within participatory and deliberative democratic processes (Seskir et al., 2023).

Wolbring (2022) highlights a small focus on considering the 'social' within quantum technology discussions—specifically in equity, diversity, and inclusion- to underscore the need for engagement with a broader range of stakeholders in democratizing quantum technologies. This gap and a greater need to make quantum technologies more understandable (Vermaas, 2017) could further hinder engagement with necessary stakeholders, lead to a failure of democratization.

**Toward Responsible Climate-Oriented Innovation**

This insight is particularly critical when considering quantum's potential role in addressing climate change. As the results indicate, quantum's potential to support climate mitigation, through applications such as carbon capture simulations or advanced materials for renewable energy, is frequently celebrated across press releases.

While quantum advances could significantly aid environmental monitoring, renewable energy optimization, and sustainable material discovery, realizing these potentials equitably will require more than technical breakthroughs. It demands governance frameworks that promote public deliberation, address access inequalities, and ensure climate-oriented quantum innovations align with broader societal needs rather than narrow commercial or geopolitical interests (Seskir et al., 2023).

Moreover, Seskir et al. (2023) noted that the militarization and strategic competition surrounding quantum technologies, particularly their incorporation into national security agendas, may hinder democratization efforts and restrict society's ability to steer their development toward the public good. In the context of climate challenges, this risk becomes pronounced: if quantum capabilities become ingrained within geopolitical competition rather than directed toward global challenges such as mitigating the impacts of climate change, the opportunity for transformative environmental impacts could be lost.

The "hype helix" model underscores that the relationship between scientific promises and national strategic visions is not linear but iterative. Expectations are created, amplified, and fed back into the system, influencing research priorities, policy agendas, and public discourse. Although hype has successfully mobilized investment and political support, it also risks fostering unrealistic expectations that could erode public trust if promised outcomes are delayed or unmet (Roberson et al., 2023).

In the context of climate change, managing these expectations becomes exceptionally crucial. Yet, as exemplified by the strategic and geopolitical themes, there is a real risk that quantum development may be steered by national security goals rather than environmental or societal benefit.

Promises that quantum technologies will significantly aid climate mitigation and adaptation must be tempered by an awareness of the field's technological uncertainties and long development timelines. Without such care, there is a real danger that quantum's potential contributions to climate action will be overstated, potentially diverting attention and resources from more immediately viable solutions.

Finally, future discourse must more explicitly address global disparities in access to quantum technologies and participation in its workforce. Without conscious efforts to prioritize ethical, security, and equity considerations, the development of quantum technologies may inadvertently exacerbate existing technological and social divides. Embedding these concerns into the governance of quantum innovation will ensure that the field contributes meaningfully to sustainable and inclusive outcomes.

**Contributions and Future Work**

This study helps contribute to the growing body of scholarship on the societal framing of emerging technologies by demonstrating how quantum is positioned within discourses of sustainability, economic development, and technological integration. It highlights the increasingly strategic nature of quantum narratives and offers a nuanced understanding of how innovation in this field is intertwined with broader societal and institutional ambitions. Our findings also support a more critical and reflective perspective on how quantum's societal role is being actively constructed across sectors.

Building on the contributions of this study within the realm of communication on climate change mitigation through quantum technology, future research must monitor the evolution of these narratives, particularly as quantum technologies move closer to widespread application. This study's thematic findings—particularly the interplay between synergy with emerging technologies and equity concerns—offer a baseline for monitoring how quantum narratives evolve as commercialization expands.

Researchers must assess whether integration with AI and other fields yields substantive environmental and societal benefits or primarily reinforces visions that fail to achieve equitable outcomes.

**Limitations**

While qualitative research methods offer rich insights into complex social narratives, they also come with several limitations that are important to acknowledge in this study of press release discourse on quantum technologies and climate change (Queirós et al., 2017). First, the process is highly time-consuming; using the Constant Comparative Method (Glaser, 1965) and applying over 8,500 codes to 100 press releases required sustained effort, iteration, and deep

Can Quantum Physics Help Fight Climate Change? A Thematic Analysis of Press Release Narratives

familiarity with the content. Alternative methods of analysis could more efficiently gain insights from far larger samples. Second, like many qualitative studies, this research lacks generalizability. The press releases analyzed reflect a specific six-month period and a purposive sample shaped by keyword filters. As such, the findings provide insight into trends but cannot be generalized to all communications about quantum technologies and their association as a solution to climate change. Furthermore, keywords such as climate change might inherently reflect its reality as a problem.

      Additionally, researcher bias is a key limitation. The interpretation of narrative themes—such as "Hope vs. Caution" or "Equity and Access"—is inherently subjective, shaped by the researcher's perspective and analytical lens. Thematic coding, even when supported by software like MAXQDA, still depends on the researcher's judgment. The study also faces challenges common to qualitative research methods, including difficulty establishing causal relationships and analyzing and synthesizing diverse data sources. Finally, because the research is grounded in specific texts and interpretive processes, it has limited replicability; another researcher might identify different themes or emphasize alternative aspects of the discourse (Queirós et al., 2017). These limitations do not diminish the value of the findings but rather highlight future considerations for research on this topic.

Can Quantum Physics Help Fight Climate Change? A Thematic Analysis of Press Release Narratives

**Conclusion**

This research reveals that quantum technologies are increasingly portrayed as powerful tools in the global response to climate change. Press releases from academia, government, and industry frame quantum as a scientific breakthrough and a strategic asset capable of enabling more efficient energy systems, driving innovation, and supporting environmental monitoring. These narratives are shaped by overlapping themes that blend technical potential with societal, economic, and geopolitical considerations.

However, while the discourse is essentially optimistic, it is also interlaced with cautionary hints of ethical risks, cybersecurity threats, and the importance of equitable access. The convergence of quantum technologies with AI and other emerging technologies underscores the need for integrated approaches to governance and innovation. As the world spotlights quantum science during the UN's International Year of Quantum in 2025, it is critical to understand how these narratives influence public expectations, policy agendas, and future investment. Ultimately, how quantum technologies are communicated today will shape their social acceptance, trajectory, and impact on climate action tomorrow.

Can Quantum Physics Help Fight Climate Change? A Thematic Analysis of Press Release Narratives